\documentstyle[emulateapj]{article}
\begin{document}
\submitted{Revised version March 6, 2000}
\title{Supernovae versus Neutron Star Mergers 
as the Major $r$-Process Sources}
\author{Y.-Z. Qian}
\affil{School of Physics and Astronomy, University of
Minnesota, Minneapolis, MN 55455; qian@physics.umn.edu}

\begin{abstract}
I show that recent observations of $r$-process abundances in metal-poor stars
are difficult to explain if neutron star mergers (NSMs) are the major 
$r$-process sources. In contrast,
such observations and meteoritic data on $^{182}$Hf and $^{129}$I in the
early solar system support a self-consistent picture of $r$-process 
enrichment by supernovae (SNe). 
While further theoretical studies of $r$-process
production and enrichment are needed for both SNe and NSMs, I emphasize two 
possible direct observational tests of the SN $r$-process model: gamma rays 
from decay of $r$-process nuclei in SN 
remnants and surface contamination of the companion by SN $r$-process ejecta
in binaries.
\end{abstract}
\keywords{Galaxy: evolution --- nuclear reactions, nucleosynthesis, 
abundances --- supernovae: general --- stars: neutron}

\section{Introduction}

All of the actinides 
and approximately half of the stable nuclei with mass numbers 
$A\gtrsim 100$ in the
solar system were produced by rapid neutron capture, the $r$-process.
While the astrophysical site of this process remains a mystery, the main
candidate environments are neutrino-heated ejecta from core-collapse
supernovae (SNe; Woosley \& Hoffman 1992; Meyer et al. 1992; 
Takahashi, Witti, \& Janka 1994; Woosley et al. 1994)
and decompressed ejecta from neutron star mergers
(NSMs; Lattimer \& Schramm 1974, 1976; Symbalisty \& Schramm 1982;
Freiburghaus, Rosswog, \& Thielemann 1999b). To be a major source for the
$r$-process, an environment must satisfy two criterions: one on reproducing
the solar $r$-process abundance pattern and the other on supplying the
total amount of $r$-process material in the present Galaxy. Using more
physical parametrizations than the previous approach 
based on constant neutron
number density and temperature (e.g., Kratz et al. 1993), several recent
studies (Hoffman, Woosley, \& Qian 1997; Meyer \& Brown 1997; Freiburghaus
et al. 1999a) derived the $r$-process conditions that can produce relative
abundance patterns with peaks at $A\sim 130$ and 195 as observed in the
solar system. At present, it seems that conditions in {\it models} of 
neutrino-heated ejecta from SNe are deficient (e.g., Qian \& Woosley 1996)
while those in {\it models} of decompressed ejecta from NSMs are promising
(e.g., Freiburghaus et al. 1999b) for an $r$-process.
However, due to the uncertainties 
in the theoretical models
(e.g., Qian \& Woosley 1996; Freiburghaus et al. 1999b), 
a reliable comparison of 
the {\it actual} conditions in these two environments 
with the $r$-process conditions cannot be
made yet to help establish or discriminate either environment
as the $r$-process
site. Furthermore, for both environments the models showed that
if the ejecta were composed of $r$-process material, then the amount 
provided by a single event combined with the estimated number
of SNe or NSMs over Galactic history would be adequate to account for
the present Galactic $r$-process inventory (e.g.,
Mathews \& Cowan 1990; Qian \& Woosley 1996; Rosswog et al. 1999).
Therefore, the above two criterions cannot readily identify SNe or
NSMs as the major $r$-process sources. 

In this Letter I discuss a phenomenological approach 
to test SNe and NSMs as the major 
$r$-process sources.
By considering mixing of the ejecta from an individual event
with the interstellar medium (ISM),
I show that observations of metal-poor 
stars are difficult to explain if NSMs are the major $r$-process sources
(\S2). I further show that a self-consistent picture of $r$-process
enrichment by SNe based on the same consideration
is supported by meteoritic data on 
$^{182}$Hf and $^{129}$I in the early solar system 
and by observations of metal-poor stars (\S3). 
To emphasize the importance of observations in establishing the major
$r$-process sources,
I conclude with a discussion of two possible direct tests of the
SN $r$-process model:
gamma rays from decay of $r$-process nuclei in
SN remnants and surface contamination of the companion by SN $r$-process
ejecta in binaries (\S4).

\section{$r$-Process Abundances and Neutron Star Mergers}

I assume that as far as the $r$-process is concerned, 
the mergering of a neutron star (NS) with a black hole (BH)
is similar to that of two neutron stars.
These two kinds of events are collectively referred to as NSMs.
The rate of such events in the Galaxy is quite uncertain. 
Here I take an average rate 
$\langle f_{\rm G}^{\rm NSM}\rangle\sim (10^4\ {\rm yr})^{-1}$ 
over Galactic history, which is at the very
high end of various estimates (e.g., Phinney 1991; Bethe \& Brown 1998;
Arzoumanian, Cordes, \& Wasserman 1999). Numerical simulations of
a NS-NS merger event by Rosswog et al. (1999) showed that a total mass
$M_{\rm ej}^{\rm NSM}\sim 4\times (10^{-3}$--$10^{-2})M_\odot$ of
decompressed NS material may be ejected. Then the grand total
from all the past NSMs over Galactic history of $t_{\rm G}\sim 10^{10}$~yr
is $M_{\rm ej}^{\rm NSM}\langle f_{\rm G}^{\rm NSM}\rangle t_{\rm G}\sim
4\times (10^3$--$10^4)M_\odot$.
This is roughly equal to the present Galactic $r$-process inventory 
$X_{\odot,r}^{\rm tot}M_{\rm G}\sim 10^4\,M_\odot$, where
$X_{\odot,r}^{\rm tot}\sim 10^{-7}$ 
is the total $r$-process mass fraction of nuclei with $A\gtrsim 100$ in the
solar system (K\"appeler, Beer, \& Wisshak 1989)
and $M_{\rm G}\sim 10^{11}\,M_\odot$ is the total Galactic mass in
stars and gas. So it seems that NSMs could be the major
$r$-process sources.

However, the amount of ejecta from a single NSM event discussed above
is too much to explain the observed $r$-process abundances in metal-poor
stars. This can be seen by considering mixing of the ejecta from each
event with the ISM. 
Rosswog et al. (1999) showed that
the total energy of the NSM ejecta is at most comparable to the SN 
explosion energy ($\sim 10^{51}$~erg). Therefore, when its 
energy/momentum is dispersed in the ISM, this ejecta can mix with at
most the same amount of material as swept up by a SN remnant,
$M_{\rm mix}\approx 3\times 10^4\,M_\odot$ (e.g., Thornton et al. 1998).
Consequently, if all of this ejecta were $r$-process material as required
to account for the present Galactic $r$-process inventory, an ISM
enriched by a single NSM event would have a total $r$-process mass
fraction
\begin{equation}
X_{r,{\rm NSM}}^{\rm tot}
=1.3\times 10^{-7}
\left({M_{\rm ej}^{\rm NSM}\over 
4\times 10^{-3}\,M_\odot}\right)\left({3\times 10^4\,M_\odot\over
M_{\rm mix}}\right).
\label{xnsm}
\end{equation}
The solar $r$-process mass fractions of elements with 
$100\lesssim A\leq 130$ and $A>130$ are 
$X_{\odot,r}^{100\lesssim A\leq 130}\approx X_{\odot,r}^{A>130}\approx
4\times 10^{-8}$ (K\"appeler et al. 1989). 
According to equation (\ref{xnsm}), 
$X_{r,{\rm NSM}}^{\rm tot}$ would be approximately equal to
$X_{\odot,r}^{\rm tot}\approx X_{\odot,r}^{100\lesssim A\leq 130} +
X_{\odot,r}^{A>130}$ even for the lowest $M_{\rm ej}^{\rm NSM}$ of interest.
Therefore, whether the NSM ejecta were composed of $r$-process elements
with $100\lesssim A\leq 130$ or $A>130$, or both groups in solar proportion,
equation (\ref{xnsm}) predicts that abundance ratios of e.g.,
Ag ($A\sim 107$) and/or Eu ($A\sim 153$) with respect to hydrogen 
in an ISM enriched by a single event would be
at least comparable to the corresponding solar $r$-process values
(Ag/H)$_{\odot,r}$ and (Eu/H)$_{\odot,r}$. 
This predicted level of $r$-process
enrichment is in disagreement with
recent observations of $r$-process abundances in metal-poor stars, as
Ag/H in stars with ${\rm [Fe/H]}\sim -1.3$ to $-2.2$ are $\sim 10$--$10^2$
times lower than (Ag/H)$_{\odot,r}$ (Crawford et al. 1998)
while Eu/H in stars with ${\rm [Fe/H]}\sim -3$ are $\sim 30$--$10^3$ lower
than (Eu/H)$_{\odot,r}$ (McWilliam et al. 1995; 
Sneden et al. 1996, 1998). 

The following interpretation of equation (\ref{xnsm}) gives some 
insights into how NSMs would explain the present Galactic
$r$-process inventory and may help appreciate why this explanation is
disfavored by observations. The Galaxy can be divided into 
$\sim 3\times 10^6$ units each
having a mass $M_{\rm mix}\approx 3\times 10^4\,M_\odot$. This division
has a physical meaning as $M_{\rm mix}$ is the maximum mass within which
the ejecta from an individual NSM event can be distributed. 
Due to the rather
low rate of NSMs in the Galaxy, on average at most one NSM event
occurred in a unit over Galactic history. Therefore, in order to
account for the present Galactic $r$-process inventory, each unit would
have to be enriched with an approximately solar $r$-process mass fraction
by a single NSM event. As shown above, this picture of $r$-process
enrichment is inconsistent with the observed $r$-process abundances in
metal-poor stars. Furthermore, due to the high rate of SNe in the Galaxy,
many SNe occurred in a unit where a single NSM event also occurred
at sometime in Galactic history (see \S4). As Fe enrichment of this unit was
provided by these SNe, stars formed at different times
in this unit would have varying
[Fe/H] but either zero or approximately solar $r$-process mass fraction
if NSMs were the major $r$-process sources. This is in disagreement
with the observed
correlation between abundances of $r$-process elements and Fe at
${\rm [Fe/H]}\gtrsim -2.5$ (Gratton \& Sneden 1994; Crawford et al. 1998;
see also McWilliam et al. 1995).

Of course, the above discussion of mixing of the NSM ejecta and the
ISM is oversimplified. For example, one could imagine that a smaller than
average fraction of $r$-process ejecta was mixed into the ISM near the
edge of a NSM remnant. In this case stars formed near the edge of the
remnant would have $r$-process mass fractions smaller than that given
by equation (\ref{xnsm}). However, one would also expect that less than
average enrichment was not unduly pervasive and a significant fraction
of the metal-poor stars would have the $r$-process enrichment indicated
by equation (\ref{xnsm}). The fact that no such stars have been observed
suggests a difficulty of the NSM $r$-process model in explaining
the observations at low metallicities. Furthermore, even if the observed 
$r$-process abundances in metal-poor stars could be attributed to pervasive 
less than average enrichment by NSMs, one would still face the other
difficulty in explaining the observed correlation between abundances of
$r$-process elements and Fe at ${\rm [Fe/H]}\gtrsim -2.5$. As Fe 
enrichment was controlled by SNe occurring at a much higher rate,
widely-varying degrees of mixing of the $r$-process ejecta in an already
existing NSM remnant with Fe produced by fresh SNe would result in large
scatter in $r$-process abundances over a broad range of [Fe/H]. This is
in disagreement with the rather early onset of the correlation between
abundances of $r$-process elements and Fe at ${\rm [Fe/H]}\approx -2.5$.
In summary, observations of metal-poor stars would be difficult to 
explain if NSMs were the major $r$-process sources.
 
\section{$r$-Process Enrichment by Supernovae}

In contrast to the case of NSMs, observations of metal-poor stars
as well as meteoritic data on $^{182}$Hf and $^{129}$I in the early solar
system support a self-consistent picture of $r$-process enrichment by SNe
(Qian \& Wasserburg 2000, hereafter QW; Wasserburg \& Qian 2000; 
hereafter WQ). 
In this picture
the ejecta from each SN event is mixed with an average mass
$M_{\rm mix}\approx 3\times 10^4\,M_\odot$ of ISM swept up by the SN
remnant (e.g., Thornton et al. 1998). It is assumed that the SN rate
per unit mass of gas $f_{\rm G}^{\rm SN}/M_{\rm gas}$
is approximately constant over Galactic history
(this seems reasonable as the star formation rate is proportional to
the gas mass). Consequently, an average ISM in the Galaxy
is enriched by newly-synthesized material from SNe at a frequency
\begin{eqnarray}
f_{\rm mix}^{\rm SN}&=&M_{\rm mix}{f_{\rm G}^{\rm SN}\over M_{\rm gas}}
=(10^7\ {\rm yr})^{-1}\nonumber\\
&\times&\left({M_{\rm mix}\over 3\times 10^4\,M_\odot}\right)
\left[{f_{\rm G}^{\rm SN}\over (30\ {\rm yr})^{-1}}\right]
\left({10^{10}\,M_\odot\over M_{\rm gas}}\right),
\label{fmix}
\end{eqnarray}
where $f_{\rm G}^{\rm SN}/M_{\rm gas}$
is estimated using quantities for
the present Galaxy. 

Meteoritic data on $^{182}$Hf (with a lifetime $\bar\tau_{182}
=1.3\times 10^7$~yr) and $^{129}$I ($\bar\tau_{129}=2.3\times 10^7$~yr)
in the early solar system
shed important light on the $r$-process and its association with SNe.
Wasserburg, Busso, \& Gallino (1996; see also QW) showed
that the abundance ratio $({^{182}{\rm Hf}}/{^{180}{\rm Hf}})_{\rm SSF}
=2.4\times 10^{-4}$ (Harper \& Jacobsen 1996; Lee \& Halliday 1995, 1998)
is consistent with common SNe injecting $^{182}$Hf into the ISM
at a high frequency $f_{\cal{H}}\approx f_{\rm mix}^{\rm SN}\sim
(10^7\ {\rm yr})^{-1}$ over a uniform production time $T_{\rm UP}\approx
10^{10}$~yr before solar system formation (SSF). However, 
the abundance ratio 
$({^{129}{\rm I}}/{^{127}{\rm I}})_{\rm SSF}=10^{-4}$ (Reynolds 1960; see
also Brazzle et al. 1999) must be accounted for by a different type of
SNe occurring at a low frequency $f_{\cal{L}}\sim f_{\cal{H}}/10\sim
(10^8\ {\rm yr})^{-1}$ (Wasserburg et al. 1996; QW). 
Therefore, the meteoritic data require at least two distinct types of
SN $r$-process events: the high-frequency 
$\cal{H}$ events producing heavy nuclei
with $A>130$ including $^{182}$Hf and the low-frequency 
$\cal{L}$ events producing
light nuclei with $A\leq 130$ including $^{129}$I. The $r$-process 
production in the SN environments associated with the $\cal{H}$ and 
$\cal{L}$ events was
discussed in some detail by Qian, Vogel, \& Wasserburg (1998a).

The above picture of $r$-process production and enrichment by SNe has clear
predictions for $r$-process abundances resulting from a single event 
(QW; WQ). For example, with an average ISM enriched by 
the $\cal{H}$
events at a frequency $f_{\cal{H}}\sim (10^7\ {\rm yr})^{-1}$, the solar
$r$-process abundances of elements with $A>130$ such as Eu
were provided by
$f_{\cal{H}}T_{\rm UP}\sim 10^3$ $\cal{H}$ events.
This requires that the Eu abundance in an ISM enriched by a single $\cal{H}$ 
event must be
\begin{equation}
\log\epsilon_{\cal{H}}({\rm Eu})=\log\epsilon_{\odot,r}({\rm Eu})
-\log(f_{\cal{H}}T_{\rm UP})\sim -2.5,
\label{epseu}
\end{equation}
where the $\log\epsilon$ notation is defined as
$\log\epsilon({\rm Eu})\equiv\log({\rm Eu/H})+12$ with Eu/H being the
abundance ratio of Eu to hydrogen, and $\log\epsilon_{\odot,r}({\rm Eu})
=0.51$ is the value for solar $r$-process Eu 
(K\"appeler et al. 1989).
Similarly, the solar $r$-process abundances of elements with
$A\leq 130$ such as Ag were provided by $f_{\cal{L}}T_{\rm UP}\sim 10^2$
$\cal{L}$ events. This requires that the Ag abundance in an ISM enriched
by a single $\cal{L}$ event must be
\begin{equation}
\log\epsilon_{\cal{L}}({\rm Ag})=\log\epsilon_{\odot,r}({\rm Ag})
-\log(f_{\cal{L}}T_{\rm UP})\sim -0.8,
\label{epsag}
\end{equation}
where $\log\epsilon_{\odot,r}({\rm Ag})=1.19$ 
(K\"appeler et al. 1989) is used.
The predictions in equations (\ref{epseu}) and (\ref{epsag})
are in good agreement with observations of very metal-poor stars which
were formed when only a small number of SNe had contributed $r$-process
elements to the ISM.
The observed $\log\epsilon({\rm Eu})$ values for stars
with ${\rm [Fe/H]}\sim -3$
range from $-2.5$ to $-0.9$ (McWilliam et al. 1995; 
Sneden et al. 1996, 1998),
which can be accounted for by $\sim 1$--40 $\cal{H}$ events with
$\log\epsilon_{\cal{H}}({\rm Eu})\sim -2.5$ from a single event
(QW; WQ).
In addition, Ag abundances at the level 
indicated by equation (\ref{epsag}) were
observed in HD 2665 and BD +37$^\circ$1458 with ${\rm [Fe/H]}\sim -2$
(Crawford et al. 1998) and in CS 22892--052 with ${\rm [Fe/H]}\sim -3$
(Cowan et al. 1999).

In the above picture of $r$-process enrichment by SNe 
the total mass of $r$-process elements 
ejected in an $\cal{H}$ event must be
$X_{\odot,r}^{A>130}M_{\rm mix}/
(f_{\cal{H}}T_{\rm UP})\sim 10^{-6}\,M_\odot$, 
while that in an $\cal{L}$ event
must be $X_{\odot,r}^{100\lesssim A\leq 130}M_{\rm mix}/
(f_{\cal{L}}T_{\rm UP})
\sim 10^{-5}\,M_\odot$.
A total $\sim 10^{-6}$--$10^{-5}\,M_\odot$ of material can be
naturally provided by the neutrino-heated ejecta 
from the proto-neutron star in a SN over a period $\sim 1$--10~s
(e.g., Qian \& Woosley 1996). However, whether this neutrino-heated
ejecta is composed of $r$-process material is yet to be shown.
The difference by a factor $\sim 10$
in the total amount of $r$-process ejecta between the $\cal{H}$ and $\cal{L}$
events has been speculated to indicate that neutrino emission and hence, 
ejection of $r$-process
material are terminated by the transition of the proto-neutron star into a
BH in the $\cal{H}$ events while 
prolonged ejection is sustained by
neutrino emission from a stable NS in the $\cal{L}$ events 
(Qian et al. 1998a).

In summary, despite the lack of a complete model for 
successful $r$-process production in SNe, there is a self-consistent picture
of $r$-process enrichment
by SNe that can account for the meteoritic data on $^{182}$Hf and $^{129}$I 
in the early solar system and 
the observed $r$-process abundances in metal-poor
stars. Furthermore, 
the observed correlation between abundances 
of $r$-process elements and
Fe at ${\rm [Fe/H]}\gtrsim -2.5$ 
(Gratton \& Sneden 1994; Crawford et al. 1998;
see also McWilliam et al. 1995) can be understood as the result of
sufficient mixing of $r$-process products and Fe from multiple SN events
in this picture. In fact,
the same picture was used by Wasserburg \& Qian (WQ)
as the basic framework to explain the dispersion in abundances
of heavy $r$-process elements such as Eu at ${\rm [Fe/H]}\sim -3$.

\section{Discussion and Conclusions}

The amount of ISM to mix with the ejecta from an individual NSM event
is limited by the total energy of the event. On the other hand, due to
the very low rate of NSMs in the Galaxy, a large amount of $r$-process
ejecta would be required from each event to account for the present
Galactic $r$-process inventory. When mixed with the ISM, this required
amount of ejecta would result in abundances of $r$-process elements
with $A\leq 130$ (such as Ag) and $A>130$ (such as Eu) that are much too
high (by factors $\sim 10$--10$^2$ for Ag and $\sim 30$--$10^3$ for Eu)
compared with the observed values in metal-poor stars. 
Furthermore, an average ISM received
the ejecta from only $\sim 1$ NSM event over Galactic history. If 
$r$-process enrichment of the ISM was provided by NSMs in this way while
Fe enrichment was provided by many SNe, 
there would be no correlation
between abundances of $r$-process elements and Fe, in disagreement
with the observed correlation at ${\rm [Fe/H]}\gtrsim -2.5$.
While the complexities in mixing of the ejecta with the ISM can affect
the above considerations in detail, it is unlikely that they can remove
the difficulty of the NSM $r$-process model in explaining the 
observations of metal-poor stars (especially the rather early onset of
the correlation between abundances of $r$-process elements and Fe at
${\rm [Fe/H]}\approx -2.5$). Nevertheless, future numerical studies of
$r$-process enrichment by NSMs accompanied by Fe enrichment by SNe
should be interesting to pursue and may give a more definitive answer.

In contrast, a self-consistent picture of 
$r$-process enrichment by SNe can be obtained by considering mixing of
the ejecta from an individual event with the ISM. Here
an average ISM is enriched in $r$-process elements with $A>130$ by
the $\cal{H}$ events at a frequency 
$f_{\cal{H}}\sim (10^7\ {\rm yr})^{-1}$ and in those with $A\leq 130$
by the $\cal{L}$ events at a frequency 
$f_{\cal{L}}\sim (10^8\ {\rm yr})^{-1}$. This picture can account for
the meteoritic data on $^{182}$Hf and $^{129}$I in the early solar
system and the observed $r$-process abundances in metal-poor stars.
Furthermore, sufficient mixing of $r$-process products and Fe from
multiple SN events in this picture would result in the observed
correlation between abundances of $r$-process elements and Fe at
${\rm [Fe/H]}\gtrsim -2.5$. The same picture was also used by
Wasserburg \& Qian (WQ) as the basic framework to explain the 
dispersion in abundances of heavy $r$-process elements such as
Eu at ${\rm [Fe/H]}\sim -3$. However, a complete model of 
$r$-process and Fe production in the $\cal{H}$ and $\cal{L}$ events
is still lacking and should be investigated by future theoretical
studies.

In discussing $r$-process enrichment by NSMs I have assumed that 
the maximum amount of ISM to mix with the ejecta from
an individual event is the same as swept up by a SN remnant. This
is because the total energy of the NSM ejecta seen in numerical
simulations (Rosswog et al. 1999) is at most comparable to the
SN explosion energy ($\sim 10^{51}$~erg). 
For given conditions of the ISM, 
the expansion/evolution of the ejecta is essentially governed by 
its total energy. The large difference in the amount of ejecta
between a NSM and a SN has no significant effect here as in both cases
the mass of the swept-up ISM soon overwhelms that of the ejecta
while the total energy remains more or less the same. 
I note that a small amount ($\lesssim 10^{-5}\,M_\odot$)
of material might be ejected
in highly-relativistic jets in
a NS-BH merger event (Janka et al. 1999). However, the total energy
of these jets is $\lesssim 10^{51}$~erg (Janka et al. 1999)
and their existence would
not increase significantly the amount of ISM that could mix with
the entire ejecta from the event. It takes $\sim 10^6$~yr 
for the energy ($\sim 10^{51}$~erg) and the associated 
momentum of the ejecta to be 
dispersed in the ISM (e.g., Thornton et al. 1998), where the
next NSM or SN would occur on a much longer timescale 
($\sim 10^{10}$~yr for NSMs and $\sim 10^{7}$~yr for SNe).
This leaves substantial time for mixing of the ejecta with the
swept-up ISM. However, the details of the mixing process
require further studies.

If, as argued here, SNe are the major sources for the $r$-process,
then there are two possible direct observational tests: gamma rays
from decay of $r$-process nuclei in SN remnants and surface 
contamination of the companion by SN $r$-process ejecta in binaries.
Qian et al. (1998b, 1999) discussed gamma-ray signals
from decay of long-lived $r$-process nuclei
(with lifetimes $\sim 10^3$--$10^5$~yr) in a nearby SN remnant and
from decay of short-lived $r$-process nuclei (with lifetimes
$\sim 1$--10~yr) produced in a Galactic SN that may occur in the
future. The nuclide $^{126}$Sn is of particular interest 
(Qian et al. 1998b) as its lifetime
($\sim 10^5$~yr) is much longer than the age ($\sim 10^4$~yr) of
the Vela SN remnant at a distance $\approx 250$~pc. 
In the picture of $r$-process enrichment by SNe discussed here
(see also QW; WQ) the solar $r$-process mass fraction of nuclei with
$A\leq 130$ resulted from $\sim 10^2$ $\cal{L}$ events (see \S3). 
So a total mass $X_{\odot,r}^{A=126}M_{\rm mix}/10^2
\sim 5\times 10^{-7}\,M_\odot$ of $^{126}$Sn nuclei are produced
in each $\cal{L}$ event, where $X_{\odot,r}^{A=126}\approx
1.6\times 10^{-9}$ (K\"appeler et al. 1989) is the solar $r$-process
mass fraction of $^{126}$Te, the stable daughter of $^{126}$Sn.
If the SN associated with the Vela remnant was an $\cal{L}$ event,
then decay of $^{126}$Sn in this remnant would produce gamma-ray
fluxes $\sim 10^{-7}\,\gamma\ {\rm cm}^{-2}\ {\rm s}^{-1}$ at
energies $E_\gamma=415$, 666, and 695 keV. Detection of these fluxes
would require future gamma-ray experiments such as the proposed 
Advanced Telescope for High Energy Nuclear Astrophysics 
(ATHENA, Kurfess 1994). As the Vela remnant contains a pulsar,
such detection would also provide evidence for the speculated 
association between $\cal{L}$ events and SNe producing neutron stars 
(Qian et al. 1998a).

The other test mentioned above takes advantage of
the occurrence of SNe in binaries. The $r$-process ejecta from
the SN would contaminate the surface of its binary companion.
Some binaries would survive the SN explosion and acquire a NS or
a BH in place of the SN progenitor. Therefore,
an ordinary star observed to be the binary companion of 
a NS or a BH might show $r$-process abundance anomalies 
on the surface. To estimate the plausible level of such anomalies,
I assume that a fraction $\sim 10^{-3}$ of the entire SN ejecta 
($\sim 10\,M_\odot$,
mostly non-$r$-process material) would be intercepted
by a low mass ($\sim 1\,M_\odot$) companion and then mixed with
$\sim 10^{-2}\,M_\odot$ of the surface material. If the SN was an
$\cal{H}$ event, $\sim 10^{-9}\,M_\odot$ of $r$-process elements
with $A>130$ would be intercepted, while for an $\cal{L}$ event
$\sim 10^{-8}\,M_\odot$ of $r$-process elements with $A\leq 130$
would be intercepted (see \S3). These quantities are to be compared with
$\sim 4\times 10^{-10}\,M_\odot$ of the corresponding $r$-process 
elements in $\sim 10^{-2}\,M_\odot$ of the surface material in
a companion star of solar $r$-process composition. So a SN in a
binary could enhance significantly the surface $r$-process 
abundances in the companion star, especially if the SN was an
$\cal{L}$ event. In view of the large overabundance of 
O, Mg, Si, and S recently observed in the companion star of a 
probable BH (Israelian et al. 1999), detection of 
$r$-process enhancement in similar binary systems seems promising.
Such detection may also test directly the speculation by
Qian et al. (1998a) that $\cal{H}$ events are associated with
SNe producing BHs while $\cal{L}$ events are associated
with SNe producing neutron stars.

\acknowledgments

I thank Petr Vogel and Jerry Wasserburg for many discussions on 
the $r$-process. I am also grateful to the referee, Friedel Thielemann,
for detailed criticisms that help improve the paper.
This work was supported in part by the Department of Energy under
grant DE-FG02-87ER40328.


\begin{references}

\reference{}
Arzoumanian, Z., Cordes, J. M., \& Wasserman, I. 1999, \apj, 520, 696

\reference{}
Bethe, H. A., \& Brown, G. E. 1998, \apj, 506, 780

\reference{}
Brazzle, R. H., Pravdivtseva, O. V., Meshik, A. P., \& Hohenberg, C. M. 1999,
\gca, 63, 739

\reference{}
Cowan, J. J., Sneden, C., Ivans, I., Burles, S., Beers, T. C., \&
Fuller, G. 1999, \baas, 31, 930

\reference{}
Crawford, J. L., Sneden, C., King, J. R., Boesgaard, A. M., 
\& Deliyannis, C. P. 1998, \aj, 116, 2489

\reference{}
Freiburghaus, C., Rembges, J.-F., Rauscher, T., Kolbe, E., Thielemann, F.-K.,
Kratz, K.-L., Pfeiffer, B., \& Cowan, J. J. 1999a, \apj, 516, 381

\reference{}
Freiburghaus, C., Rosswog, S., \& Thielemann, F.-K. 1999b, \apjl, 525, L121

\reference{}
Gratton, R. G., \& Sneden, C. 1994, \aap, 287, 927

\reference{}
Harper, C. L., \& Jacobsen, S. B. 1996, \gca, 60, 1131

\reference{}
Hoffman, R. D., Woosley, S. E., \& Qian, Y.-Z. 1997, \apj, 482, 951

\reference{}
Israelian, G., Rebolo, R., Basri, G., Casares, J., \& Martin, E. L. 1999,
\nat, 401, 142

\reference{}
Janka, H.-Th., Eberl, Th., Ruffert, M., \& Fryer, C. L. 1999, \apjl, 527, L39

\reference{}
K\"appeler, F., Beer, H., \& Wisshak, K. 1989, Rep. Prog. Phys., 52, 945

\reference{}
Kratz, K.-L., Bitouzet, J.-P., Thielemann, F.-K., M\"oller, P., \& 
Pfeiffer, B. 1993, \apj, 403, 216

\reference{}
Kurfess, J. D. 1994, ATHENA Mission Proposal, NASA New Mission Concepts 
in Astrophysics

\reference{}
Lattimer, J. M., \& Schramm, D. N. 1974, \apjl, 192, L145

\reference{}
Lattimer, J. M., \& Schramm, D. N. 1976, \apj, 210, 549

\reference{}
Lee, D.-C., \& Halliday, A. N. 1995,  Nature, 378, 771

\reference{}
Lee, D.-C., \& Halliday, A. N. 1998,  
Lunar and Planetary Science Conference, 29, 1416 

\reference{}
Mathews, G. J., \& Cowan, J. J. 1990, \nat, 345, 491

\reference{}
McWilliam A., Preston, G. W., Sneden, C., \& Searle, L. 1995, \aj, 109, 2757

\reference{}
Meyer, B. S., \& Brown, J. S. 1997, \apjs, 112, 199

\reference{}
Meyer, B. S., Howard, W. M., Mathews, G. J., Woosley, S. E., \&
Hoffman, R. D. 1992, \apj, 399, 656

\reference{}
Qian, Y.-Z., Vogel, P., \& Wasserburg, G. J. 1998a, \apj, 494, 285

\reference{}
Qian, Y.-Z., Vogel, P., \& Wasserburg, G. J. 1998b, \apj, 506, 868

\reference{}
Qian, Y.-Z., Vogel, P., \& Wasserburg, G. J. 1999, \apj, 524, 213

\reference{}
Qian, Y.-Z., \& Wasserburg, G. J. 2000, Phys. Rep., in press (QW)

\reference{}
Qian, Y.-Z., \& Woosley, S. E. 1996, \apj, 471, 331

\reference{}
Reynolds, J. H. 1960, Phys. Rev. Lett., 4, 8

\reference{}
Rosswog, S., Liebend\"orfer, M., Thielemann, F.-K., Davies, M. B., Benz, W.,
\& Piran, T. 1999, \aap, 341, 499

\reference{}
Sneden, C., Cowan, J. J., Burris, D. L., \& Truran, J. W. 1998, 
\apj, 496, 235

\reference{}
Sneden, C., McWilliam, A., Preston, G. W., Cowan, J. J., Burris,
D. L., \& Armosky, B. J. 1996, \apj, 467, 819

\reference{}
Symbalisty, E., \& Schramm, D. N. 1982, Astrophys. Lett., 22, 143

\reference{}
Takahashi, K., Witti, J., \& Janka, H.-Th. 1994, \aap, 286, 857

\reference{}
Thornton, K., Gaudlitz, M., Janka, H.-Th., \& Steinmetz, M. 1998, 
\apj, 500, 95

\reference{}
Wasserburg, G. J., Busso, M., \& Gallino, R. 1996, \apjl, 466, L109

\reference{}
Wasserburg, G. J., \& Qian, Y.-Z. 2000, \apjl, 529, L21 (WQ)

\reference{}
Woosley, S. E., \& Hoffman, R. D. 1992, \apj, 395, 202

\reference{}
Woosley, S. E., Wilson, J. R., Mathews, G. J., Hoffman, R. D., \&
Meyer, B. S. 1994, \apj, 433, 229

\end{references}
\end{document}